\documentclass[aps,twocolumn,superscriptaddress,showpacs]{revtex4}

\usepackage[dvips]{graphicx}
\newcommand{\nc}{\newcommand}           % new command
\nc{\vc}[1]     {\mbox{\boldmath $#1$}} % boldmath(vector)
\nc{\mydraft}	{\setlength{\topmargin}{-1.5cm}}
\mydraft

\begin{document}

\title{Mirror symmetry breaking in He isotopes and their mirror nuclei}

\author{Takayuki Myo\footnote{myo@ge.oit.ac.jp}}
\affiliation{General Education, Faculty of Engineering, Osaka Institute of Technology, Osaka, Osaka 535-8585, Japan}
\affiliation{Research Center for Nuclear Physics (RCNP), Osaka University, Ibaraki 567-0047, Japan}

\author{Kiyoshi Kat\=o\footnote{kato@nucl.sci.hokudai.ac.jp}}
\affiliation{Nuclear Data Centre, Faculty of Science, Hokkaido University, Sapporo 060-0810, Japan}
                     
\date{\today}

\begin{abstract}%
We study the mirror symmetry breaking of $^6$He-$^6$Be and $^8$He-$^8$C using the $^4$He + $XN$~($X$ = 2, 4) cluster model.
The many-body resonances are treated for the correct boundary condition using the complex scaling method.
We find that the ground state radius of $^8$C is larger than that of $^8$He due to the Coulomb repulsion in $^8$C. 
On the other hand, the $0^+_2$ resonances of the two nuclei exhibit the inverse relation; the $^8$C radius is smaller than the $^8$He radius. 
This is due to the Coulomb barrier of the valence protons around the $^4$He cluster core in $^8$C, 
which breaks the mirror symmetry of the radius in the two nuclei. 
A similar variation in the radius is obtained in the mirror nuclei, $^6$He and $^6$Be.
A very large spatial extension of valence nucleons is observed in the $0^+_2$ states of $^8$He and $^8$C.
This property is related to the dominance of the $(p_{3/2})^2(p_{1/2})^2$ configuration for four valence nucleons,
which is understood from the reduction in the strength of the couplings to other configurations by involving the spatially extended components of valence nucleons.
\end{abstract}


\maketitle 

\section{Introduction} \label{sec:intro}

Unstable nuclei have often been observed in unbound states beyond the particle thresholds due to the weak binding nature of valence nucleons.
Resonance spectroscopy of unbound states in unstable nuclei has been developed using radioactive-beam experiments \cite{tanihata13}.
In addition to the energies and decay widths of resonances, information on their configurations and spatial properties is important in understanding the structures of unstable nuclei.
In proton-rich and neutron-rich nuclei, the correlations between valence nucleons and between a valence nucleon and a core nucleus 
lead to the exotic nuclear properties in resonances as well as in the weakly bound states that are currently beyond our standard understanding of stable nuclei.
The comparison of the structures between proton-rich and neutron-rich nuclei is related to the mirror symmetry in unstable nuclei with a large isospin.

The results of certain experiments on proton-rich $^8$C have recently been reported \cite{charity11}. The $^8$C nucleus is an unbound system beyond the proton drip line.
Currently, only the ground state is observed in $^8$C. This resonance is located at 2 MeV above the $^6$Be + $2p$ threshold energy and close to the $^7$B + $p$ threshold \cite{charity11}.
The excited states of $^8$C can decay not only to a two-body $^7$B + $p$ channel, 
but also to the many-body channels of $^6$Be + $2p$, $^5$Li + 3$p$ and $^4$He + 4$p$, in which $^5$Li, $^6$Be, and $^7$B are also unbound systems.
Large decay widths for multi-particle decays make it difficult to identify the excited states of $^8$C experimentally.

The mirror nucleus of $^8$C is $^8$He with isospin $T=2$. Thus far, many experiments on $^8$He have been reported \cite{tanihata92,skaza07,golovkov09}.
Its ground state is weakly bound, and it has a neutron skin structure consisting of four valence neutrons around $^4$He with a separation energy of 3.1 MeV.
As regards the excited states of $^8$He, most of these states are observed above the $^4$He + 4$n$ threshold energy.
Therefore, the resonances of $^8$He can decay into many channels including $^7$He + $n$, $^6$He + 2$n$, $^5$He + 3$n$, and $^4$He + 4$n$, similar to $^8$C.
The multi-particle decays of $^8$He are related to the Borromean nature of $^6$He, which decays into $^4$He + $2n$ with a small excitation energy.
With this background information on He isotopes and their mirror nuclei, we remark that most of the states appearing in these nuclei can exist as many-body resonances.
It is important to understand the characteristics of these resonances as many-body states beyond the simple two-body resonance, 
considering the couplings with the continuum states of various decay channels into subsystems. 
The proper treatment of boundary conditions for the resonant and continuum states is inevitable to obtain reliable results of the resonance properties.

The comparison of the structures of proton- and neutron-rich nuclei is interesting from the viewpoint of mirror symmetry.
In this study we compare the structures of $^8$He and $^8$C and also those of $^6$He and $^6$Be.
It is interesting to examine the effect of the Coulomb interaction on the mirror symmetry of these nuclei.
The structures of resonances and weakly bound states are generally influenced by the open channels of particle emissions.
Hence, the mirror symmetry of unstable nuclei can be related to the coupling behavior between the open channels and the continuum states.

In the theoretical analysis to investigate the unbound states of $^8$He and $^8$C, it is necessary to describe the $^4$He + 4$n$/4$p$ five-body unbound states.
In this analysis, we employ the cluster-orbital shell model (COSM) \cite{suzuki88,masui06,myo077,myo097,myo10,myo117,myo128,myo13,horiuchi12} that has been successfully used to describe the $^4$He+4$N$ five-body system.
In COSM, we adopt single particle wave functions of non-restricted radial forms, which are optimized by solving the many-body Schr\"odinger equation. 
In addition, the effects of all the open channels are taken into account explicitly in the COSM, and therefore we can treat the many-body decaying phenomena.
In our previous works \cite{myo10,myo128,myo13}, we have successfully described the He isotopes and their mirror, proton-rich nuclei.
We have obtained many-body resonances using the complex scaling method (CSM) \cite{ABC1,ABC2,aoyama06} under the correct boundary conditions for all decay channels. 
In CSM, the resonant wave functions are described using the $L^2$ basis functions without any approximation.
Using the combined COSM and CSM we describe the many-body resonances and also the decaying process of the states.
The application of CSM to nuclear physics has been extensively carried out for resonance spectroscopy, transition strengths into unbound states, and nuclear reaction problems 
\cite{myo097,myo10,aoyama06,myo01,myo0711,kruppa07,kikuchi13,ogata13}. 

In the previous analyses \cite{myo117}, we have discussed the mirror symmetry of $^7$He and $^7$B with the $^4$He + 3$N$ model.
It is found that the mirror symmetry is broken in their ground states where 
the mixing of $^6$Be($2^+_1$) in $^7$B is larger than that of $^6$He($2^+_1$) in $^7$He.
This result concerns the relative energies between the states of $A$ = $7$ nuclei and the channels of the ($2^+_1$ state of $A$ = $6$ nuclei)+$N$. 
In Refs. \cite{myo128,myo13}, we investigated the five-body resonances of $^8$C with the $^4$He + $4p$ model.
We found that the configurations of $^8$He and $^8$C are similar; the ground $0^+$ states are mainly described 
by the $(p_{3/2})^4$ configuration for four valence nucleons as a sub-closed nature,
and the excited $0^+_2$ states are strongly dominated by the 2p2h configuration of $(p_{3/2})^2(p_{1/2})^2$.

In this study, we proceed with our analysis of $^8$He and $^8$C and concentrate on the spatial properties of the $0^+$ states of the two nuclei.
Recently, the dineutron-cluster model \cite{kobayashi13} has been used to demonstrate the possibility of viewing dineutron clustering in $^8$He($0^+_2$) as like $^4$He-core plus two dineutrons.
This model is based on the bound-state approximation.
On the other hand, our results for the $0^+_2$ states of $^8$He and $^8$C show a single configuration of $(p_{3/2})^2(p_{1/2})^2$ of valence nucleons with the correct treatment for resonances.
It is important to clarify the reason for the single configuration of the $0^+_2$ states 
in relation to the boundary condition of resonances and also to the spatial extension of valence nucleons because the $0^+_2$ states are resonances decaying into five-body constituents.
In addition, we compare the radii of $^8$He and $^8$C to understand the effect of the Coulomb interaction on the spatial distribution of the valence protons in $^8$C,
which was partially carried out in the previous study \cite{myo128}.
In this study we examine the mirror symmetry in the spatial properties of the two nuclei.
The Coulomb interaction provides a repulsive effect on the system, which is naively expected to enlarge the radius of the system.
In fact, we have shown that the ground state radius of $^8$C is larger than that of $^8$He \cite{myo128,myo13}.
We perform the same analysis for the $0^+_2$ resonances of the two nuclei. 
The comparison between mirror nuclei is also discussed for the $A$ = $6$ nuclei, $^6$He and $^6$Be using the $^4$He + $2N$ model.

This paper is organized as follows.
In Sect.~\ref{sec:model}, we explain the COSM wave function and the complex scaling method.
In Sect.~\ref{sec:result}, we present the results of the mirror symmetry relations of $^8$He-$^8$C and that of $^6$He-$^6$Be.
A summary is provided in Sect.~\ref{sec:summary}.
\section{Complex-scaled cluster orbital shell model} \label{sec:model}

We use the COSM of the $^4$He + 4$N$ system for $^8$He and $^8$C, and the $^4$He + 2$N$ system for $^6$He and $^6$Be.
The Hamiltonian is the same as that used in the previous studies~\cite{myo117,myo128},
\begin{eqnarray}
	H
&=&	\sum_{i=1}^{X+1}{t_i} - T_G + \sum_{i=1}^{X} V^{\alpha N}_i + \sum_{i<j}^{X} V^{NN}_{ij}
    \\
&=&	\sum_{i=1}^{X} \left[ \frac{\vc{p}^2_i}{2\mu} + V^{\alpha N}_i \right] + \sum_{i<j}^{X} \left[ \frac{\vc{p}_i\cdot \vc{p}_j}{4m} + V^{NN}_{ij} \right] ,
    \label{eq:Ham}
\end{eqnarray}
where $X$ denotes the valence nucleon number around $^4$He, and the mass number is given by 4 + $X$.
The terms $t_i$ and $T_G$ denote the kinetic energies of each particle ($^4$He and $N$) and of the center of mass of the total system, respectively.
The operator $\vc{p}_i$ denotes the relative momentum between $^4$He and $N$.
The reduced mass $\mu$ is $4m/5$ described using the nucleon mass $m$.
The $^4$He-$N$ interaction $V^{\alpha N}$ is given by the microscopic KKNN potential \cite{aoyama06,kanada79} for the nuclear part,
in which the tensor correlation of $^4$He is renormalized.
For the Coulomb part, we use the folded Coulomb potential using the density of $^4$He as the $(0s)^4$ configuration.
The $NN$ interaction $V^{NN}$ is given by the Minnesota potential \cite{tang78} for the nuclear part in addition to the point Coulomb interaction between protons.
These interactions reproduce the low-energy scatterings of the $^4$He-$N$ and the $N$-$N$ systems, respectively.

For the wave function of $^4$He, we assume a $(0s)^4$ configuration of a harmonic oscillator wave function with length parameter of 1.4 fm.
We expand the relative wave functions of the $^4$He + $XN$ system ($X=2, 4$) using the COSM basis states \cite{suzuki88}.
In the COSM, the total wave function $\Psi^J$ with spin $J$ is represented by the superposition of the configuration $\Psi^J_c$ as
\begin{eqnarray}
    \Psi^J
&=& \sum_c C^J_c \Psi^J_c,
    \label{WF0}
    \\
    \Psi^J_c
&=& \prod_{i=1}^{X} a^\dagger_{\alpha_i}|0\rangle, 
    \label{WF1}
\end{eqnarray}
where the vacuum $|0\rangle$ represents the $^4$He ground state.
The creation operator $a^\dagger_{\alpha}$ corresponds to the single-particle state of a valence nucleon above $^4$He
with the quantum number $\alpha=\{n,\ell,j\}$ in the $jj$-coupling scheme.
The index $n$ represents the different radial components. 
In this study, the valence nucleons are assumed to be protons only or neutrons only.  
The index $c$ represents a set of $\alpha_i$ values for valence nucleons as $c=\{\alpha_1,\ldots,\alpha_{X}\}$. 
We obtain a summation over the available configurations in Eq.~(\ref{WF0}) with a total spin $J$.
The amplitudes $\{C_c^J\}$ in Eq.~(\ref{WF0}) are determined variationally 
with respect to the total wave function $\Psi^J$ by the diagonalization of the Hamiltonian matrix elements.

%%%%%%%%%%%%%%%%%%%%%%%%%%%%%
\begin{figure}[t]
\centering
\includegraphics[width=6.8cm,clip]{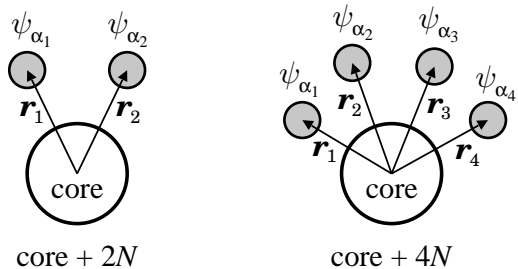}
\caption{Sets of the spatial coordinates in COSM.}
\label{fig:COSM}
\end{figure}
%%%%%%%%%%%%%%%%%%%%%%%%%%%%%
 
The coordinate representation of the single-particle state for $a^\dagger_{\alpha}$ is given as 
$\psi_{\alpha}(\vc{r})$ as a function of the relative coordinate $\vc{r}$ between $^4$He and a valence nucleon, as shown in Fig.~\ref{fig:COSM}.
We employ a sufficient number of bases to expand $\psi_\alpha(\vc{r})$ for the description of the spatial extension of valence nucleons, in which the $\psi_\alpha(\vc{r})$ terms are normalized.
In this model, the radial part of $\psi_\alpha(\vc{r})$ is expanded with the Gaussian basis functions for each orbit \cite{masui06,aoyama06} as
\begin{eqnarray}
    \psi_\alpha(\vc{r})
&=& \sum_{k=1}^{N_{\ell j}} d^k_{\alpha}\ \phi_{\ell j}^k(\vc{r},b_{\ell j}^k),
    \label{WFR}
    \\
    \phi_{\ell j}^k(\vc{r},b_{\ell j}^k)
&=& {N}\, r^{\ell} {\rm e}^{-(r/b_{\ell j}^k)^2/2} [Y_{\ell}(\hat{\vc{r}}),\chi^\sigma_{1/2}]_{j},
    \label{Gauss}
	\\
    \langle \psi_\alpha|\psi_{\alpha'} \rangle 
&=& \delta_{\alpha,\alpha'}.
    \label{Gauss2}
\end{eqnarray}
The index $k$ corresponds to the Gaussian basis function with the length parameter $b_{\ell j}^k$.
The normalization factor of the basis and basis number are given by ${N}$ and $N_{\ell j}$, respectively, and $k=1,\ldots,N_{\ell j}$.
The coefficients $\{d^k_{\alpha}\}$ in Eq.~(\ref{WFR}) are determined using the Gram-Schmidt orthonormalization.
Hence, the basis states $\psi_\alpha(\vc{r})$ are orthogonal to each other, as indicated by Eq.~(\ref{Gauss2}).
The radial bases $\psi_\alpha(\vc{r})$ are distinguished by the index $n$, which is indexed up to the basis number $N_{\ell j}$.
The same method using Gaussian basis functions for a single-particle state is employed in the tensor-optimized shell model \cite{myo0711,myo09,myo11,myo12}.
The Pauli principle between a valence nucleon and $^4$He is treated using the orthogonality condition model \cite{saito77}, 
in which the single-particle state $\psi_{\alpha}(\vc{r})$ is imposed to be orthogonal to the $0s$ state occupied by $^4$He.
The length parameters $b_{\ell j}^k$ are chosen in geometric progression \cite{aoyama06}.
We use at most 17 Gaussian basis functions by setting $b_{\ell j}^k$ to range from 0.2 fm to around 40 fm typically.
The large value of $b_{\ell j}^k$ plays an important role in accounting for the continuum state of a valence nucleon.
To obtain the Hamiltonian matrix elements of multi-nucleon systems in the COSM, we employ the $j$-scheme technique of the shell-model calculation.

For the model space of the COSM, the angular momenta of the single-particle states are taken as $\ell\le 2$ \cite{myo10,myo128}. 
We fit the two-neutron separation energy of $^6$He to the experiment of 0.975 MeV by taking the 173.7 MeV of the repulsive strength of the Minnesota potential from the original value of 200 MeV \cite{tang78}.

%%%%%%%%%%%%%%%%%%%%%%%%%%%%%%%%%%%%%%%%%%%%%
%\subsection{Complex scaling method (CSM)}

We briefly review the CSM to describe the resonances and nonresonant continuum states.
In the CSM, we transform the relative coordinates as $U(\theta) \vc{r}_i = \vc{r}_i\, {\rm e}^{i\theta}$ for $i=1,\ldots,X$, as shown in Fig. \ref{fig:COSM}, 
where $U(\theta)$ denotes the operator for transformation with a scaling angle $\theta$.
The Hamiltonian in Eq.~(\ref{eq:Ham}) is transformed into the complex-scaled Hamiltonian $H_\theta=U(\theta) H U(\theta)^{-1}$, and the corresponding complex-scaled Schr\"odinger equation is given as
\begin{eqnarray}
	H_\theta\Psi^J_\theta
&=&     E\Psi^J_\theta .
	\label{eq:eigen}
\end{eqnarray}
The eigenstates $\Psi^J_\theta=U(\theta) \Psi^J$ are obtained by solving the eigenvalue problem of $H_\theta$.
In the CSM, all the energy eigenvalues $E$ of the bound and unbound states are located on a complex energy plane, according to the ABC theorem \cite{ABC1,ABC2}.
In this theorem, it is proved that the boundary condition of the resonances is transformed into damping behavior at the asymptotic region.
This condition is the same as that of the bound states.
For a finite value of $\theta$, every Riemann branch cut starting from different particle thresholds is commonly rotated down by $2\theta$ in the complex energy plane.
In contrast, bound states and resonances are obtainable as stationary solutions with respect to $\theta$.
We can identify the pole of resonances as Gamow states with complex eigenvalues, $E=E_r-{\rm i}\Gamma/2$, where $E_r$ and $\Gamma$ denote the resonance energies and decay widths, respectively. 
In the wave function, the $\theta$-dependence is included in the amplitudes in Eq.~(\ref{WF0}) as $\{C_c^J(\theta)\}$.
The optimal angle $\theta$ is determined to locate the stationary point of each resonance in the complex energy plane \cite{aoyama06}.

In the CSM, resonances are precisely described as eigenstates expanded in terms of $L^2$ basis functions such as Gaussian basis functions.
It is noted that the amplitude $\{C_c^J(\theta)\}$ of resonances becomes a complex number, 
while the resonances are normalized to obey the relation $\sum_{c} \left(C_c^J(\theta)\right)^2=1$ \cite{myo077,homma97,berggren96}.
For amplitudes, when the imaginary part is a relatively small value, it is reasonable to discuss the physical meaning of the real part of the amplitudes.
In the present study, we also discuss the radius of resonances being a complex number.
When the radius exhibits an imaginary part relatively smaller than the real one,
we consider that the real part of the radius is regarded as a reference providing the information on the spatial size of the resonances.
Recently, the complex radius has been used to estimate the size of hadronic resonances such as those observed in the $\bar{K}N$ system \cite{dote13,sekihara13}.

The Hermitian product is not applied due to the bi-orthogonal eigenstates of the non-Hermitian $H_\theta$ \cite{berggren68,homma97,myo98}.
The matrix elements of a physical quantity $\hat{O}$ for resonant states are calculated using the amplitudes $\{C_c^J(\theta)\}$ and are independent of the angle $\theta$,
as evidenced from the following relation
\begin{eqnarray}
   \langle\tilde{\Psi}|\hat{O}|\Psi\rangle
&=&\langle\tilde{\Psi} U(\theta)^{-1}|U(\theta)\hat{O}U(\theta)^{-1}|U(\theta)\Psi\rangle
   \nonumber\\
&=&\langle\tilde{\Psi}_\theta|\hat{O}_\theta|\Psi_\theta\rangle 
   \nonumber\\
&=&\sum_{c\, c'} C^J_c(\theta) C^J_{c'}(\theta) \langle\tilde{\Psi}_c^J |\hat{O}_\theta|\Psi_{c'}^J \rangle .
   \label{eq:ME}
\end{eqnarray}
The properties of the resonances are uniquely determined as the Gamow states \cite{aoyama06,romo68}.
In the CSM, it is not necessary to obtain the original solution of resonances by using the inverse transformation of the complex scaling, which is the so-called back-rotation. 
This is because the matrix elements of the resonances are uniquely obtained in Eq.~(\ref{eq:ME}).
\section{Results} \label{sec:result}

\subsection{Energy spectra}

%%%%%%%%%%%%%%%%%%%%%%%%%%%%%%
% 6He/Src05.4/
% 6Be/Src05/
\begin{table}[t]
\caption{Energy eigenvalues of the $0^+_{1,2}$ states of $^6$He and $^6$Be measured from the $^4$He + $N$ + $N$ threshold with the experimental values. Units are in MeV.}
\label{ene6}
\centering
\small
\begin{ruledtabular}
\begin{tabular}{c|ccccc}
        & \multicolumn{2}{c }{$^6$He}  & & \multicolumn{2}{c}{$^6$Be}    \\ 
        & Energy          &  $\Gamma$  & & Energy         &  $\Gamma$    \\ \hline
$0^+_1$ & $-$0.975        &  ---       & &  $1.383$       &  0.041       \\
 exp.   & $-$0.975        &  ---       & &  $1.370$       &  0.092       \\ \hline
$0^+_2$ & 3.88            &  8.76      & &  $5.95$        &  11.21       \\
\end{tabular}
\end{ruledtabular}
\end{table}
%%%%%%%%%%%%%%%%%%%%%%%%%%%%%%

%%%%%%%%%%%%%%%%%%%%%%%%%%%%%%
% 8He/Src01.6/
% C8/Src01/Data01/16_0975_5, 17_0.900
\begin{table}[t]
\caption{Energy eigenvalues of the $0^+_{1,2}$ states of $^8$He and $^8$C measured from the $^4$He + $N$ + $N$ + $N$ + $N$ threshold with the experimental values \cite{charity11}. 
Units are in MeV.}
\label{ene8}
\centering
\small
\begin{ruledtabular}
\begin{tabular}{c|ccccc}
        & \multicolumn{2}{c}{$^8$He} && \multicolumn{2}{c}{$^8$C}       \\ \hline
        & Energy        &  $\Gamma$   && Energy       &  $\Gamma$        \\
$0^+_1$ & $-$3.22       &  ---        && 3.32         &  0.072           \\
 exp.   & $-$3.11       &  ---        && 3.449(30)    &  0.130(50)       \\ \hline
$0^+_2$ & 3.07          &  3.19       && 8.88             &  6.64        \\
\end{tabular}
\end{ruledtabular}
\end{table}
%%%%%%%%%%%%%%%%%%%%%%%%%%%%%%

We discuss the structures of the $0^+$ states of $^8$He and $^8$C, and also those of $^6$He and $^6$Be.
The energy eigenvalues of these states have been obtained in a previous analysis \cite{myo128}, and here, we discuss the mirror symmetry of these nuclei.
We summarize the energy properties in Tables~\ref{ene6} and \ref{ene8}, which values are measured from the $^4$He + $N$ + $N$(+ $N$ + $N$) thresholds.
For $^8$He and $^8$C, the $0^+_{2}$ states are obtained as five-body resonances.
The systematic energy spectra for He isotopes and their mirror nuclei have been presented in previous studies \cite{myo10,myo128}.
The results well reproduce the observed energy levels and predict several resonances.
The matter and charge radii of $^6$He and $^8$He obtained in the COSM also agree with the observations \cite{myo10}.
This agreement indicates that the spatial distributions of the valence neutrons are described well to explain the neutron halo and skin structures in the weakly bound states.
In the following section, we discuss the various radii values corresponding to the ground and resonant states to investigate the spatial properties of valence nucleons in relation to the mirror symmetry.

%%%%%%%%%%%%%%%%%%%%%%%%%%%%%%%%%%%%%%%%%%%%%%%%%%%%%%%%%%%%
\subsection{Effect of spatial extension of valence nucleons in $^8$He and $^8$C}

We discuss the effect of the spatial extension of valence nucleons on the configuration mixing of the single-particle states in $^8$He and $^8$C.
This analysis is important to understand the reason for the dominance of a single configuration in the $0^+_2$ resonances of the two nuclei.
In the COSM, we expand the wave function of the valence nucleons using the Gaussian basis functions given in Eq.~(\ref{Gauss}) with a length parameters $\{b^k_{\ell j} \}$.
The spatial extension is considered in a superposition of the Gaussian basis functions with the large value of $b^k_{\ell j}$.
To investigate the effect of spatial extension, we restrict the motion of valence nucleons in which the relative distance between $^4$He and a valence nucleon becomes smaller than 6 fm.
The distance of 6 fm is determined to retain the properties of the configuration mixing of the $^8$He ground state.
Since $^8$He has a bound state, the four valence neutrons of $^8$He are considered to be distributed not very far from $^4$He.
The restriction for the spatial extension of the COSM wave functions is placed with the maximum lengths of $\{b^k_{\ell j}\}$ in Eq.~(\ref{Gauss}) for a valence nucleon.
The maximum lengths of $\{b^k_{\ell j}\}$ are determined for the Gaussian basis functions so as to satisfy the matrix elements of the relative distance between a nucleon and $^4$He (Fig.~\ref{fig:COSM}) to be 6 fm.
In this treatment, the energy of $^8$He is obtained as $-3.14$ MeV as measured from the $^4$He + $4n$ threshold, which is still close to the value of $-3.22$ MeV obtained without the restriction \cite{myo10}.

The result of the configuration mixing with the spatial restriction for valence nucleons is listed in Table~\ref{comp8_1c} for the ground states of $^8$He and $^8$C.
We list the dominant configurations with their squared amplitudes $(C^J_c)^2$ as given in Eq. (\ref{WF0}).   
The states of $^8$He and $^8$C show similar values of mixing, in which the $(p_{3/2})^4$ configuration commonly dominates the total wave functions.
The $^8$C state is described under the bound-state approximation in a manner similar to that for $^8$He, because the spatial restriction does not treat the boundary condition for resonances correctly.
For comparison, the results without the restriction are listed in Table~\ref{comp8_1} \cite{myo128}.
For $^8$He, it is found that the mixing values of each configuration do not strongly depend on the restriction of the spatial extension.
For $^8$C, the mixing values with the restriction (Table~\ref{comp8_1c}) are also close to the real parts of the values without the restriction (Table \ref{comp8_1}).
These results indicate that the spatial extension does not have a significant effect in the ground states of $^8$He and $^8$C.

%%%%%%%%%%%%%%%%%%%%%%%%%%%%%%
\begin{table}[t]
\caption{Squared amplitudes $(C^J_c)^2$ of the ground states of $^8$He and $^8$C with spatial restriction of valence nucleons.}
\label{comp8_1c}
\centering
\small
\begin{ruledtabular}
\begin{tabular}{c|ccc}
Configuration              &  $^8$He($0^+_1$) & $^8$C($0^+_1$) \\ \hline
 $(p_{3/2})^4$             &  0.858           & 0.854          \\
 $(p_{3/2})^2(p_{1/2})^2$  &  0.069           & 0.071          \\
 $(p_{3/2})^2(1s_{1/2})^2$ &  0.006           & 0.010          \\
 $(p_{3/2})^2(d_{3/2})^2$  &  0.008           & 0.008          \\
 $(p_{3/2})^2(d_{5/2})^2$  &  0.043           & 0.041          \\
 other 2p2h                &  0.011           & 0.010          \\
\end{tabular}
\end{ruledtabular}
\end{table}
%%%%%%%%%%%%%%%%%%%%%%%%%%%%%%

%%%%%%%%%%%%%%%%%%%%%%%%%%%%%%
\begin{table}[t]
\caption{Squared amplitudes $(C^J_c)^2$ of the ground states of $^8$He and $^8$C without spatial restriction.}
\label{comp8_1}
\centering
\small
\begin{ruledtabular}
\begin{tabular}{c|ccc}
Configuration              & $^8$He($0^+_1$) & $^8$C($0^+_1$)  \\  \hline
 $(p_{3/2})^4$             & 0.860           & $0.878-0.005{\rm i}$  \\
 $(p_{3/2})^2(p_{1/2})^2$  & 0.069           & $0.057+0.001{\rm i}$  \\
 $(p_{3/2})^2(1s_{1/2})^2$ & 0.006           & $0.010+0.003{\rm i}$  \\
 $(p_{3/2})^2(d_{3/2})^2$  & 0.008           & $0.007+0.000{\rm i}$  \\
 $(p_{3/2})^2(d_{5/2})^2$  & 0.042           & $0.037+0.000{\rm i}$  \\
 other 2p2h                & 0.011           & $0.008+0.000{\rm i}$  \\
\end{tabular}
\end{ruledtabular}
\end{table}
%%%%%%%%%%%%%%%%%%%%%%%%%%%%%%

%%%%%%%%%%%%%%%%%%%%%%%%%%%%%%
\begin{table}[t]
\caption{Squared amplitudes $(C^J_c)^2$ of the $0^+_2$ states of $^8$He and $^8$C with spatial restriction of valence nucleons.}
\label{comp8_2c}
\centering
\small
\begin{ruledtabular}
\begin{tabular}{c|ccc}
Configuration             & $^8$He($0^+_2$) & $^8$C($0^+_2$) \\  \hline
$(p_{3/2})^4$             & 0.069           & 0.051          \\
$(p_{3/2})^2(p_{1/2})^2$  & 0.806           & 0.796          \\
$(p_{3/2})^2(1s_{1/2})^2$ & 0.042           & 0.086          \\
$(p_{3/2})^2(d_{3/2})^2$  & 0.018           & 0.013          \\
$(p_{3/2})^2(d_{5/2})^2$  & 0.007           & 0.003          \\
other 2p2h                & 0.003           & 0.0005         \\
\end{tabular}
\end{ruledtabular}
\end{table}
%%%%%%%%%%%%%%%%%%%%%%%%%%%%%%

%%%%%%%%%%%%%%%%%%%%%%%%%%%%%%
\begin{table}[t]
\caption{Squared amplitudes $(C^J_c)^2$ of the $0^+_2$ states of $^8$He and $^8$C without spatial restriction.}
\label{comp8_2}
\centering
\small
\begin{ruledtabular}
\begin{tabular}{c|ccc}
Configuration             &  $^8$He($0^+_2$) & $^8$C($0^+_2$)  \\  \hline
$(p_{3/2})^4$             &~~$0.020-0.009{\rm i}$  &~~$ 0.044+0.007{\rm i}$ \\
$(p_{3/2})^2(p_{1/2})^2$  &~~$0.969-0.011{\rm i}$  &~~$ 0.934-0.012{\rm i}$ \\
$(p_{3/2})^2(1s_{1/2})^2$ & $-0.010-0.001{\rm i}$  & $-0.001+0.000{\rm i}$ \\
$(p_{3/2})^2(d_{3/2})^2$  &~~$0.018+0.022{\rm i}$  &~~$ 0.020+0.003{\rm i}$ \\
$(p_{3/2})^2(d_{5/2})^2$  &~~$0.002+0.000{\rm i}$  &~~$ 0.002+0.001{\rm i}$ \\
\end{tabular}
\end{ruledtabular}
\end{table}
%%%%%%%%%%%%%%%%%%%%%%%%%%%%%%

In a manner similar to the above discussion, we discuss the effect of spatial extension on the configuration mixing of the $0^+_2$ states of $^8$He and $^8$C.
In the previous analysis \cite{myo128}, it was obtained that the $(p_{3/2})^2(p_{1/2})^2$ configuration commonly dominates the total wave functions of $^8$He and $^8$C.
We restrict the spatial extension in the $0^+_2$ states in the same manner as done for the ground states. 
The results are listed in Table \ref{comp8_2c}.
We find that the configuration mixing values of the $0^+_2$ states of $^8$He and $^8$C are very similar.
The dominant configurations are retained as $(p_{3/2})^2(p_{1/2})^2$ with about 0.8 of the squared amplitudes in both nuclei.
The residual components are distributed among other configurations such as $(p_{3/2})^4$ and $(p_{3/2})^2(1s_{1/2})^2$.
As regards the results without the restriction, the dominant configurations of the valence neutron/protons are listed in Table \ref{comp8_2},
in which a nearly pure $(p_{3/2})^2(p_{1/2})^2$ configuration is obtained with very large squared amplitudes of about 0.93$-$0.97 for the real parts.
The differences in the results between Tables \ref{comp8_2c} and \ref{comp8_2} indicate that the spatial extension of the valence nucleons plays an important role 
in reducing the coupling strengths between the configurations in the $0^+_2$ states, which is not observed in the ground states of $^8$He and $^8$C.
The reduction in the coupling strength is related to the motion of valence nucleons.
Without the spatial extension, the four valence nucleons cannot be distributed widely.
In this situation, the strengths of the couplings between the different configurations generally increase 
because the amplitudes of the wave function are confined in the interaction region. 
When the spatial extension is realized in the $0^+_2$ states of $^8$He and $^8$C,
their wave functions can be distributed widely and most of the amplitudes of valence nucleons penetrate from the interaction region.
As a result, the coupling strengths between the configurations decrease and the single configuration of $(p_{3/2})^2(p_{1/2})^2$ survives dominantly in the $0^+_2$ states.

It is noted that the spatial extension of valence nucleons can also affect the diagonal energies of the single-particle configurations,
particularly with the higher orbits, because of the release of the kinetic energies of the configurations by the spatial extension of nucleons.
This effect is considered to enhance the configuration mixing because the diagonal energies of the configurations approach each other,
as opposed to the case of a reduction in the coupling matrix elements between the configurations.
The present calculation of $^8$He and $^8$C includes both effects simultaneously, and the obtained results indicate that the effect of the reduction in coupling is fairly strong.

The dineutron model calculation \cite{kobayashi13} has recently been used to determine the two-dineutron enhanced states in $^8$He($0^+_2$) under the bound-state approximation.
In general, the dineutron is a kind of spatial localization of two neutrons, such as a cluster, and it corresponds to the configuration mixing of various single-particle states in the mean field picture.
In fact, for $^6$He, the mixing of various orbits of the two valence neutrons enhances the dineutron formation \cite{aoyama06,aoyama95,hagino05}.
In the present calculations of $^8$He and $^8$C, when the spatial extension is restricted within the bound-state approximation, 
the configuration mixing in the $0^+_2$ states increase in strength in comparison with the resonance solution of the CSM.
This feature in the bound-state approximation can be related to the enhancement of the dineutron/diproton (dinucleon) components.
However, the $0^+_2$ states in the CSM calculations for $^8$He and $^8$C are described as the five-body resonances and are dominated by the single $(p_{3/2})^2(p_{1/2})^2$ configuration of four valence nucleons.
This result indicates that the states in the CSM calculations can have the different properties from the dinucleon enhanced ones.

%%%%%%%%%%%%%%%%%%%%%%%%%%%%%%
\subsection{Spatial properties of $^8$He and $^8$C}

It is important to understand the roles of the Coulomb interaction on the spatial properties of $^8$C in comparison with $^8$He.
This is related to the examination of the mirror symmetry in the two nuclei.
For this purpose, we investigate the various radii sizes of $^8$He and $^8$C, which reflect the motion of valence nucleons.
It is shown that the effect of the motion of valence nucleons on the configuration mixing depends on the state.
It is interesting to discuss the case of the radius. 
We noted that the radius of Gamow resonances is obtained as a finite complex number, 
because Gamow states have complex eigen-energies and complex amplitudes.
In the present analyses, most of the radii of resonances exhibit imaginary parts that are relatively smaller than the real ones, similar to case of the squared amplitudes 
listed in Tables~\ref{comp8_1} and \ref{comp8_2}.
Hence, we discuss the spatial size of resonances using the real part of the complex radii.
The explicit form of the operator for the various radii in the COSM is provided in Ref. \cite{suzuki88}, in which the case of He isotopes is given.

The results of the radii of the $0^+_{2}$ states in $^8$He and $^8$C are obtained as listed in Table \ref{tab:radius8} in addition to the $0^+_1$ results obtained in the previous study \cite{myo128,myo13}.
We calculate the matter ($R_{\rm m}$), proton ($R_p$), neutron ($R_n$), and charge ($R_{\rm ch}$) radii,
and the mean relative distances between $^4$He and a single valence nucleon ($r_{{\rm c}\mbox{-}N}$) 
and between $^4$He and the center of mass of the four valence nucleons ($r_{{\rm c}\mbox{-}4N}$).
For the $0^+_1$ states, the matter radius of $^8$C is larger than that of $^8$He by about 12\% for the real part.
A similar calculation for the ground states of $^8$He and $^6$He with a $^4$He core was performed using the Gamow shell model approach \cite{papadimitriou11}
in which the ground-state properties obtained for two nuclei agree with our calculation \cite{myo10}. 
The neutron-neutron correlation between two nuclei has also been investigated.
The most recent experimental value of the charge radius of $^8$He($0^+_1$) is reported as 1.959(16) fm \cite{brodeur12}, which is close to our result of 1.92 fm.

%%%%%%%%%%%%%%%%%%%%%%%%%%%%%%
\begin{table}[t]  
\caption{Various radii of the $0^+_{1,2}$ states of $^8$He and $^8$C in units of fm.}
\label{tab:radius8}
\centering
\small
\begin{ruledtabular}
\begin{tabular}{c|ccccc}
                        & $^8$He($0^+_1$) & $^8$C($0^+_1$)     & & $^8$He($0^+_2$) & $^8$C($0^+_2$)  \\ \hline
$R_{\rm m}$             & 2.52            &$2.81-0.08{\rm i}$  & & 7.56 + 2.04${\rm i}$  & 4.87 + 0.13${\rm i}$  \\
$R_p$                   & 1.80            &$3.06-0.10{\rm i}$  & & 3.15 + 0.69${\rm i}$  & 5.46 + 0.15${\rm i}$  \\
$R_n$                   & 2.72            &$1.90-0.01{\rm i}$  & & 8.53 + 2.32${\rm i}$  & 2.36 + 0.05${\rm i}$  \\
$R_{\rm ch}$            & 1.92            &$3.18-0.09{\rm i}$  & & 3.22 + 0.67${\rm i}$  & 5.53 + 0.15${\rm i}$  \\
$r_{{\rm c}\mbox{-}N}$  & 3.55            &$4.05-0.12{\rm i}$  & & 11.03+ 3.11${\rm i}$  & 7.21 + 0.21${\rm i}$  \\ % Sq[[sum<r^2(C-N)>]/4]
$r_{{\rm c}\mbox{-}4N}$ & 2.05            &$2.36-0.03{\rm i}$  & & 5.60 + 1.55${\rm i}$  & 3.68 + 0.13${\rm i}$  \\
\end{tabular}
\end{ruledtabular}
\end{table}
%%%%%%%%%%%%%%%%%%%%%%%%%%%%%%

%%%%%%%%%%%%%%%%%%%%%%%%%%%%%%
\begin{figure}[t]
\centering
\includegraphics[width=6.8cm,clip]{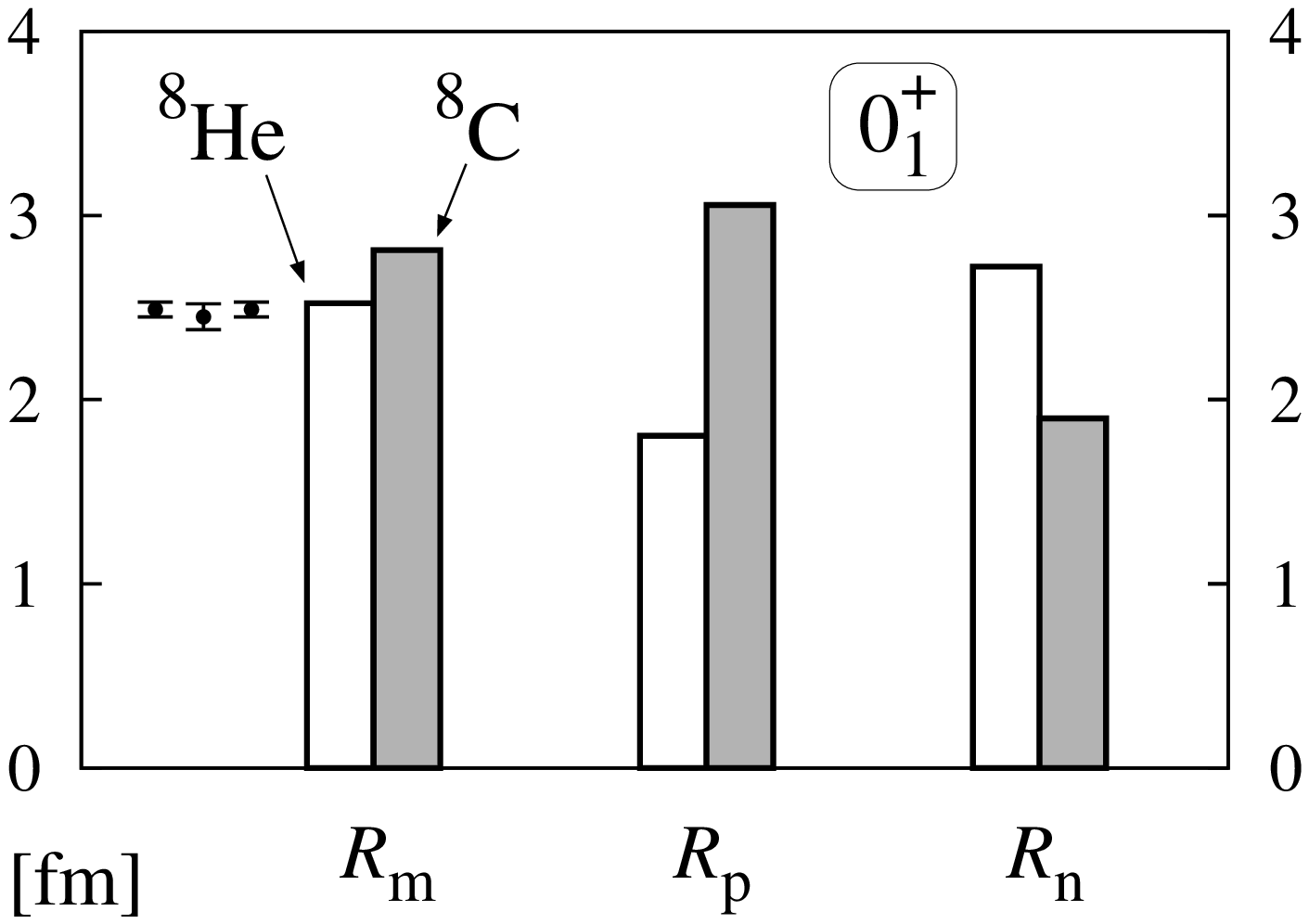}\vspace*{0.8cm}
\includegraphics[width=6.8cm,clip]{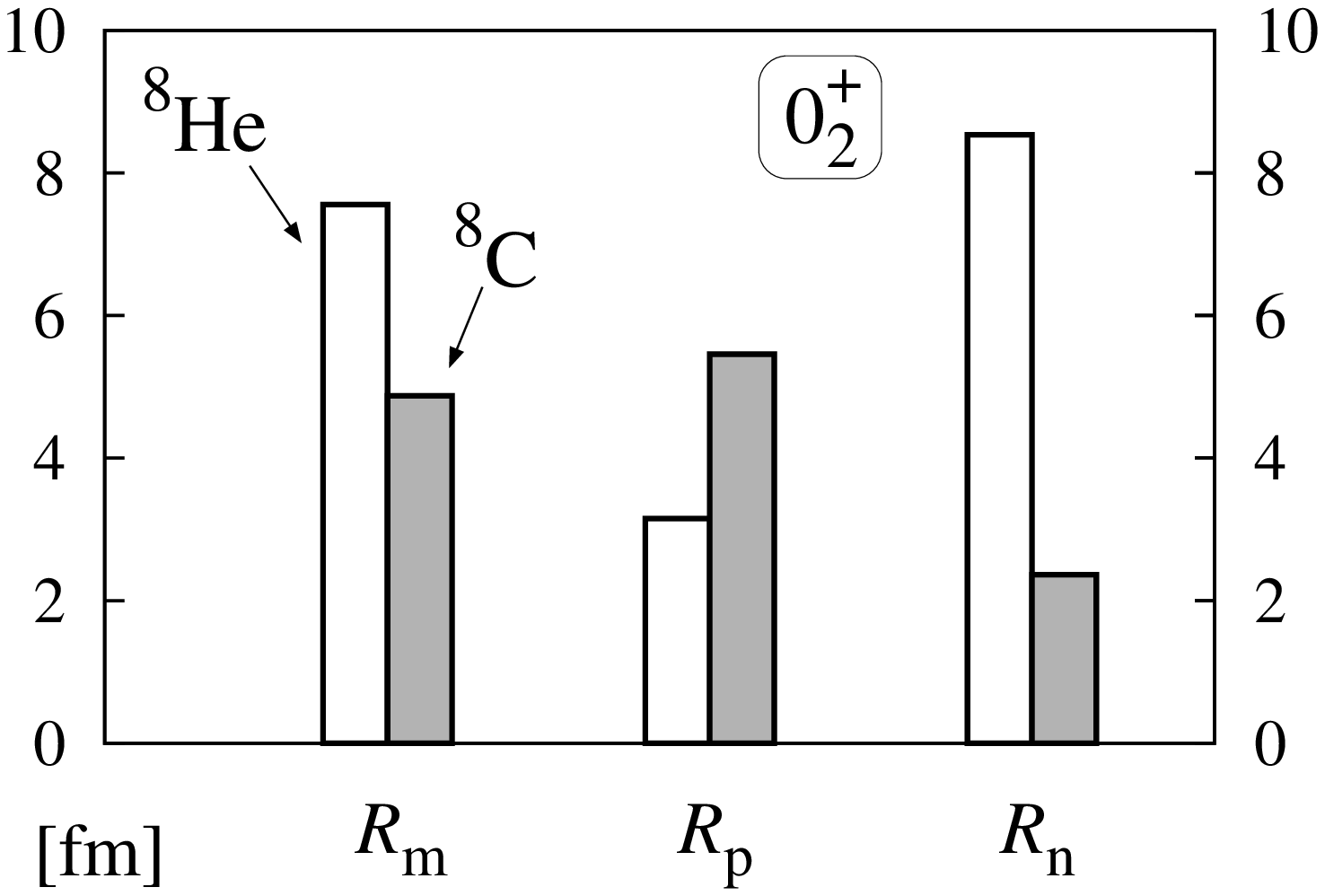}
\caption{Real parts of matter, proton, and neutron radii of the ground states (upper panel) and the $0^+_2$ states (lower panel) of $^8$He and $^8$C in units of fm. 
Circles with error bars indicate experimental data of the matter radius of $^8$He \cite{tanihata92,alkazov97,kiselev05}.}
\label{fig:radius8}
\end{figure}
%%%%%%%%%%%%%%%%%%%%%%%%%%%%%%

For $0^+_2$ states, all the values of radii are complex numbers because of the resonances in the two nuclei.
The imaginary parts are fairly large for $^8$He($0^+_2$), but still smaller than the real ones.
The values for $^8$C exhibit small imaginary parts and the bound-state properties still survive in this state,
in spite of the higher energy of 8.88 MeV from the lowest threshold.
In the two nuclei, the various radii of the $0^+_2$ states are larger than the corresponding values of their ground $0^+_1$ states.
This is consistent with the significant effect of the spatial extension of valence nucleons in the $0^+_2$ states, as discussed already.

It is interesting that the matter radius of $^8$C is smaller than that of $^8$He in the $0^+_2$ states.
The proton radius of $^8$C can be compared with the neutron radius of $^8$He.
The results indicate that the proton radius of $^8$C is smaller than the neutron radius of $^8$He.
This relation is opposite to that observed for the ground states of $^8$He and $^8$C.
To understand this difference clearly, we plot the real parts of the matter, proton, and neutron radii for $0^+_{1.2}$ in Fig. \ref{fig:radius8}.
For $^8$He($0^+_2$), the observed large matter radius originates from the large neutron radius. 
For $^8$C($0^+_2$), the large matter radius is due to the large proton radius, which is smaller than the neutron radius of $^8$He.

We conclude that the relation of the spatial properties between $^8$He and $^8$C depends on the states, which can be explained in terms of the Coulomb interaction.
The Coulomb interaction acts repulsively, thereby shifting the entire energy of $^8$C upward with respect to the $^8$He energy.
In the ground state of $^8$C, this repulsion extends the distances between $^4$He and a valence proton and between valence protons.
On the other hand, the Coulomb interaction makes the barrier above the particle threshold in $^8$C and 
the $0^+_2$ resonance is affected by this barrier, the effect of which prevents the wave function of valence protons of $^8$C from extending spatially.  
In $^8$He, there is no Coulomb barrier for the four valence neutrons and the neutrons can extend to a large distance in the resonances. 
This role of the Coulomb interaction leads to the radius of $^8$C($0^+_2$) being smaller than that of $^8$He($0^+_2$).

For the ground states of $^8$He and $^8$C, two states have different boundary conditions for the bound and resonant states. 
This difference can affect the radius in addition to the Coulomb repulsion.
Naively, the unbound condition on $^8$C is considered to enhance the $^8$C radius in comparison with the case of the bound state.
In the present calculation, the effect of the boundary condition on the radius is automatically included along with the Coulomb repulsion.
The decomposition of the two effects, while being meaningful, may be difficult to achieve.

\subsection{Spatial properties of $^6$He and $^6$Be}

For comparison with the $A$ = $8$ nuclei,
it is interesting to investigate the effect of the Coulomb interaction in $^6$He and $^6$Be, in which the spatial distributions of the two valence nucleons are involved.
Table~\ref{tab:radius6} lists the various radii of $^6$He and $^6$Be for the $0^+_{1,2}$ states.
In these states, only the ground state of $^6$He is bound, and the other $0^+$ states are three-body resonances located above the $^4$He + $N$ + $N$ threshold.
In Table~\ref{tab:radius6}, we list the matter ($R_{\rm m}$), proton ($R_p$), neutron ($R_n$), charge ($R_{\rm ch}$) radii,
the relative distances between $^4$He and a valence nucleon ($r_{{\rm c}\mbox{-}N}$),
between valence nucleons ($r_{NN}$), and between the $^4$He core and the center of mass of the two valence nucleons ($r_{{\rm c}\mbox{-}2N}$).
As regards the charge radius of $^6$He($0^+_1$), the latest experimental value is reported as 2.060(8) fm \cite{brodeur12}, which is close to our result of 2.01 fm.

It is found that the radii values for $^6$Be($0^+_1$) are mostly real, and therefore we use the real parts to estimate the radius.
The matter radius of $^6$Be($0^+_1$) is larger than that of $^6$He($0^+_1$) by 18\%, which percentage is larger than the $A$ = $8$ ($0^+_1$) case for which the corresponding value is 12\%.

%%%%%%%%%%%%%%%%%%%%%%%%%%%%%%
\begin{table}[t]  
\caption{Various radii of the $0^+_{1,2}$ states of $^6$He and $^6$Be in units of fm.}
\label{tab:radius6}
\centering
\small
\begin{ruledtabular}
\begin{tabular}{c|ccccc}
\noalign{\hrule height 0.5pt}
                        & $^6$He($0^+_1$) &  $^6$Be($0^+_1$)      & & $^6$He($0^+_2$) & $^6$Be($0^+_2$) \\ \hline 
$R_{\rm m}$             & 2.37            &2.80 + 0.17${\rm i}$ & & 3.94 + 4.12${\rm i}$  & 2.88 + 1.93${\rm i}$  \\
$R_p$                   & 1.82            &3.13 + 0.20${\rm i}$ & & 2.24 + 1.71${\rm i}$  & 3.31 + 2.33${\rm i}$  \\
$R_n$                   & 2.60            &1.96 + 0.08${\rm i}$ & & 4.57 + 4.91${\rm i}$  & 1.74 + 0.69${\rm i}$  \\
$R_{\rm ch}$            & 2.01            &3.25 + 0.21${\rm i}$ & & 2.34 + 1.63${\rm i}$  & 3.39 + 1.93${\rm i}$  \\
$r_{NN}$                & 4.82            &6.06 + 0.35${\rm i}$ & & 9.38 +10.97${\rm i}$  & 7.40 + 5.10${\rm i}$  \\
$r_{c\mbox{-}N}$        & 3.96            &4.90 + 0.40${\rm i}$ & & 7.46 + 8.06${\rm i}$  & 5.10 + 3.97${\rm i}$  \\ % Sq[[sum<r^2(C-N)>]/2 ]
$r_{{\rm c}\mbox{-}2N}$ & 3.15            &3.85 + 0.37${\rm i}$ & & 5.82 + 5.91${\rm i}$  & 3.53 + 3.07${\rm i}$  \\
\end{tabular}
\end{ruledtabular}
\end{table}
%%%%%%%%%%%%%%%%%%%%%%%%%%%%%%

%%%%%%%%%%%%%%%%%%%%%%%%%%%%%%
\begin{figure}[t]
\centering
\includegraphics[width=6.8cm,clip]{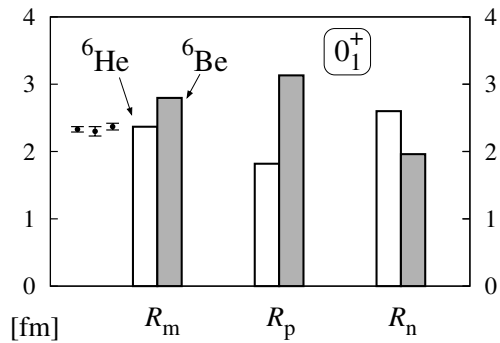}\vspace*{0.8cm}
\includegraphics[width=6.8cm,clip]{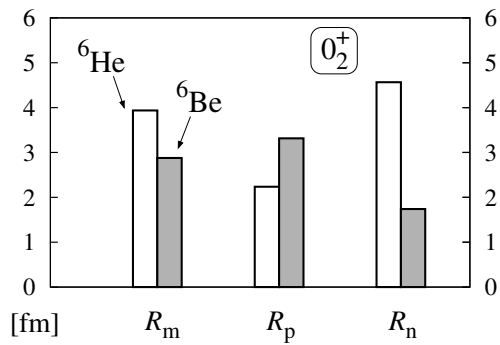}
\caption{Real parts of matter, proton, and neutron radii of the ground states (upper panel) and the $0^+_2$ states (lower panel) of $^6$Be and $^6$He in units of fm. 
Circles with error bars indicate the experimental data of the matter radius of $^6$He \cite{tanihata92,alkazov97,kiselev05}.}
\label{fig:radius6}
\end{figure}
%%%%%%%%%%%%%%%%%%%%%%%%%%%%%%

As regards the radii of the $0^+_2$ states in Table \ref{tab:radius6}, the imaginary parts of the radii increase to values larger than the real parts for certain neutron-radii related values for $^6$He.
This indicates that the coupling to the continuum states is larger in the $0^+_2$ states of the $A$ = $6$ nuclei than in the case of their ground states and also the $A$ = $8$ cases.
In a manner similar to the case of the $A$ = $8$ nuclei, we examine the mirror symmetry of the spatial properties of the $0^+_2$ state in $A$ = $6$ systems by studying the real parts of the various radii.
It is found that in the $0^+_2$ state, the $^6$Be radii values are smaller than the $^6$He radii values, which is a common feature observed in the $A$ = $8$ systems.
We also plot the real parts of the matter, proton, and neutron radii (Fig.~\ref{fig:radius6}) to examine the differences in the two nuclei.
In the ground state, the proton radius of $^6$Be is larger than the neutron radius of $^6$He due to the Coulomb repulsion from the two valence protons of $^6$Be. 
In the $0^+_2$ state, the proton radius of $^6$Be is smaller than the neutron radius of $^6$He due to the Coulomb barrier effect in $^6$Be. 
These results are due to a situation similar to that observed in the case of $^8$He and $^8$C.

From the analyses of the radius of the $0^+_{1,2}$ states for $A$ = $6$ and 8 nuclei,
we conclude that the Coulomb interaction breaks the mirror symmetry as regards the spatial properties of neutron-rich and proton-rich nuclei. 
This breaking depends on the states; spatial enhancement occurs in the $0^+_1$ states due to the Coulomb repulsion,
and the shrinkage effect influences the $0^+_2$ resonances due to the Coulomb barrier.
The latter effect on the radius can also be seen in the $3\alpha$ Hoyle state in $^{12}$C, which is located immediately above the $3\alpha$ threshold energy (Y. Funaki, private communication).
This resonance is affected by the Coulomb barrier of $\alpha$ particles, which stabilizes the Hoyle state with a fairly narrow decay width \cite{yamada04}.
There have recently been several interesting results concerning the radii of resonances in terms of both experimental \cite{danilov09} and theoretical aspects \cite{myo117,myo128,funaki05}
An effect similar to the Coulomb interaction on the stabilization of the resonances has also been reported in Ref. \cite{aoyama97}, 
in which the $s$- and $p$-wave resonances are investigated in the mirror nuclei of $^{10}$Li and $^{10}$N using the core + $N$ model.
In $^{10}$Li, the $s$-wave state exists as a virtual state. On the other hand, the corresponding mirror state in $^{10}$N transforms into a resonant state because of the presence of the Coulomb barrier.

\section{Summary} \label{sec:summary}

In summary, we investigated the spatial properties of the mirror nuclei $^8$He-$^8$C, and $^6$He-$^6$Be using the $^4$He + $XN$~($X$=2, 4) cluster model.
The boundary condition for resonances is accurately treated using the complex scaling method. 
We found that the relation of the radii between $^8$He and $^8$C is different in their ground $0^+$ states and the $0^+_2$ states.
In the ground states, the Coulomb interaction in $^8$C repulsively enlarges the radius to a value larger than that of $^8$He.
In the $0^+_2$ states, which are five-body resonances in both nuclei, 
the radius of $^8$C is smaller than that of $^8$He due to the Coulomb barrier generated from four valence protons with the $^4$He core in $^8$C. 
For $^8$He there is no Coulomb barrier, which enhances the spatial distribution of the valence neutrons in the resonance.
These roles of the Coulomb interaction break the mirror symmetry of the radius in $^8$He and $^8$C.
A similar set of variations in the radius is observed in $^6$He and $^6$Be.

We also investigated the properties of the configuration mixing of the single-particle states in the $0^+_2$ resonances of $^8$He and $^8$C.
For this purpose, we focused on the effect of the spatial extension of valence nucleons on the states.
The spatial extension of the valence nucleons plays a role in reducing the strength of the couplings between the configurations in the $0^+_2$ resonances.
This leads to the dominance of the single $(p_{3/2})^2(p_{1/2})^2$ configuration in the $0^+_2$ resonances of $^8$He and $^8$C.

\section*{Acknowledgments}
This work was supported by a Grant-in-Aid for Young Scientists from the Japan Society for the Promotion of Science (Grant No. 24740175).
Numerical calculations were performed on a supercomputer (NEC SX9) at RCNP, Osaka University.

\section*{References}
%%%%%%%%%%%%%%%%%%%%%%%%%%%%%%%%%%%%%%%%%%%%%%%%%%%%%%%%%%%%%
\def\JL#1#2#3#4{ {{\rm #1}} \textbf{#2}, #4 (#3)}  % Physical Review
\nc{\PR}[3]     {\JL{Phys. Rev.}{#1}{#2}{#3}}
\nc{\PRC}[3]    {\JL{Phys. Rev.~C}{#1}{#2}{#3}}
\nc{\PRA}[3]    {\JL{Phys. Rev.~A}{#1}{#2}{#3}}
\nc{\PRL}[3]    {\JL{Phys. Rev. Lett.}{#1}{#2}{#3}}
\nc{\NP}[3]     {\JL{Nucl. Phys.}{#1}{#2}{#3}}
\nc{\NPA}[3]    {\JL{Nucl. Phys.}{A#1}{#2}{#3}}
\nc{\PL}[3]     {\JL{Phys. Lett.}{#1}{#2}{#3}}
\nc{\PLB}[3]    {\JL{Phys. Lett.~B}{#1}{#2}{#3}}
\nc{\PTP}[3]    {\JL{Prog. Theor. Phys.}{#1}{#2}{#3}}
\nc{\PTPS}[3]   {\JL{Prog. Theor. Phys. Suppl.}{#1}{#2}{#3}}
\nc{\PRep}[3]   {\JL{Phys. Rep.}{#1}{#2}{#3}}
\nc{\JP}[3]     {\JL{J. of Phys.}{#1}{#2}{#3}}
\nc{\PPNP}[3]   {\JL{Prog.\ Part.\ Nucl.\ Phys.}{#1}{#2}{#3}}
%%%%%%%%%%%%%%%%%%%%%%%%%%%%%%%%%%%%%%%%%%%%%%%%%%%%%%%%%%%%%

\end{document}